\documentclass[10pt]{article}

\newcommand{\lsim}{\raise0.3ex\hbox{$<$}\kern-0.75em{\lower0.65ex\hbox{$\sim$}}}
\newcommand{\gsim}{\raise0.3ex\hbox{$>$}\kern-0.75em{\lower0.65ex\hbox{$\sim$}}}

\usepackage{epsfig} \usepackage{notconf}

\begin{document}

\setcounter{page}{127}

\title{Quasar environments at $0.5\leq z \leq0.8$}
\author{M. Wold$^1$, M. Lacy$^2$, P.B. Lilje$^3$, S. Serjeant$^4$\\
\newline
    $^1$Stockholm Observatory \\ $^2$IGPP, Lawrence Livermore National
    Labs and University of California, Davis \\ $^3$Institute of
    Theoretical Astrophysics, University of Oslo\\ $^4$Astrophysics
    Group, Imperial College London}

\maketitle

\begin{abstract}
Over the past few years, we have been collecting data with the Nordic
Optical Telescope (NOT) on the galaxy environments around active
galactic nuclei (AGN). Here we present some results from a sample
of 21 radio-loud and 20 radio-quiet quasars in the redshift range $0.5
\leq z \leq 0.8$.

We find a few quasars in very rich environments, perhaps as rich as
Abell class 1--2 clusters, but more often the quasars seem to prefer
groups and poorer clusters.  We also find that on average
the galaxy environments around radio-loud and radio-quiet quasars are
indistinguishable, consistent with the findings that both powerful
radio-loud and radio-quiet quasars appear to be hosted by luminous
galaxies with luminosities 
higher than the break in the luminosity function (Dunlop et al.\
1993; Taylor et al.\ 1996). Comparing the galaxy richnesses in the
radio-loud quasar fields with quasar fields in the literature, we find
a weak, but significant, correlation between quasar radio luminosity
and environmental richness.
\end{abstract}

\section{Introduction}

The differences and similarities between radio-loud and radio-quiet
quasars (hereafter RLQs and RQQs, respectively) have kept astronomers
busy for a long time.  At essentially all wavelength ranges, except at
radio wavelengths, the appearance of radio-loud and radio-quiet
quasars is similar. RQQs appear compact with a weak radio component
coinciding with the optical quasar nucleus, whereas RLQs have extended
lobes of radio emission with hotspots at the outer edges of the radio
structure. The radio-emitting lobes are being fed by powerful jets
emerging from a bright, central core.  RQQs can also have jet-like
structures (Blundell \& Beasley 1998), although with bulk kinetic
powers $\sim 10^{3}$ times lower than for RLQs (Miller, Rawlings \&
Saunders 1993). This suggests that both quasar types have
jet-producing central engines, but that the efficiency of the jet
production mechanism is very different in the two cases.
 
One way to learn more about how the two quasar types are related is to
study their host galaxies and their galactic environment. 
A longstanding belief that RQQs are hosted by spiral
galaxies and RLQs by ellipticals is now being questioned. Recent
studies (Dunlop et al.\ 1993; Taylor et al.\ 1996; McLure et al.\
1999; Hughes et al.\ 2000) have found that powerful quasars at 
$z \gsim 0.5$, both RLQs and RQQs, seem to exist in galaxies above the
break in the luminosity function at $L^{*}$, but a clear picture has
still not emerged. Some studies claim a high fraction of disk
morphologies amongst the radio-quiets (e.g.\ Percival et al.\ 2000),
whilst others suggest that nearly all quasars are in giant ellipticals
(e.g.\ McLure et al.\ 1999).

The galaxy environment on scales larger than the host galaxy is also
interesting. It may provide clues about quasar formation and evolution, 
since a period of quasar activity may be 
triggered by interactions and mergers (e.g.\ Stockton \& MacKenty
1983; Ellingson, Green \& Yee 1991).  Also, a comparison of the galaxy
environments around different types of AGN may help constrain the
so-called `Unified Models', e.g.\ the orientation-dependent unified
scheme where a RLQ and a radio galaxy are believed to be intrinsically
the same type of object, but viewed at different orientations to the
line of sight (Barthel 1989).  In the orientation-dependent unified
scheme for RLQs and radio galaxies, one expects to find that the galaxy
environments around the two are the same.

Radio galaxies and RLQs at $0.5 \leq z \leq 0.8$ are often found 
in environments with richer than average galaxy density (Yee \&
Green 1987; Hill \& Lilly 1991; Ellingson, Yee \& Green 1991; Wold et
al.\ 2000), from groups of galaxies and poorer
clusters to Abell Class 1 clusters or richer. This is perhaps not
surprising since giant elliptical galaxies frequently reside in the
centres of galaxy clusters. But what is the galaxy environment like
for RQQs in this redshift range, and how does it compare to the
environment around RLQs with comparable AGN luminosities?

In order to make a meaningful comparison one needs a method to
quantify the galaxy environment that takes into account the depth of
the survey, the angular coverage and also corrects for foreground and
background galaxies. One such parameter that has been much used when
quantifying galaxy environments around AGN is the amplitude of the
spatial cross-correlation function, the `clustering amplitude'. 
Yee \& L{\'o}pez-Cruz (1999) find that the clustering amplitude
is a robust estimator of galaxy richness in clusters.

The first attempt at comparing the environments around radio-loud and
radio-quiet quasars by using the clustering amplitude was made by Yee
\& Green (1984). They found only a marginal difference between RLQ and
RQQ fields, but later Yee \& Green (1987) did deeper imaging of the
same RQQ fields, and added more RLQ fields to the study, this time
finding an even smaller difference, but they were unable
to draw any firm conclusions due to the small number (seven) of RQQ
fields.

The problem was later addressed by Ellingson et al.\ (1991b), who also
used the clustering amplitude to quantify the environments around a
sample of 32 RLQs and 33 RQQs at $0.3 < z < 0.6$.  They found
the RLQs to exist in richer than average galaxy environment, and
frequently also in clusters as rich as Abell class 1. The RQQs, on the
other hand, were found to be much less frequently situated in rich
galaxy environments, suggesting that the two quasar types may be
physically different objects. Since this study, there has not been
much work aimed at comparing the Mpc-scale environments of the two
quasar populations at moderate redshifts using the clustering
amplitude.

We have therefore undertaken a study using the NOT to collect data on
the fields around two samples of $0.5 \leq z \leq 0.8$ RLQs and RQQs
that are matched in redshift and AGN luminosity.  The redshift range
was chosen to extend previous work which went up to $z\sim 0.6$, to as
high a redshift as possible consistent with keeping the redshifted
4000 {\AA} break shortward of the $I$-band. We selected quasars
randomly from complete flux-limited samples spanning a wide
range in both optical and radio luminosity (for the RLQs). By extending
the redshift range of previous studies, whilst maintaining the
luminosity range, we are able to better disentangle trends in the
environmental richness due to cosmic evolution from those due to radio
and/or optical luminosity. Our assumed cosmology has $H_{0}=50$
km\,s$^{-1}$\,Mpc$^{-1}$, $\Omega_{0}=1$ and $\Lambda=0$.

\section{The quasar samples}

The RLQ sample was selected from two different radio-optical flux limited
surveys, the Molonglo/APM Quasar Survey (Serjeant 1996; Maddox et
al., in prep.; Serjeant et al., in prep.)  and the 7C quasar survey
(Riley et al.\ 1999) and consists of 21 radio-loud steep-spectrum
quasars with redshifts $0.5 \leq z \leq 0.82$ covering a wide radio
luminosity range of $23.8\leq \log\left(L_{\rm 408 MHz}/{\rm W Hz}^{-1}
{\rm sr}^{-1}\right) \leq26.7$.

The 20 RQQs were selected from three optical surveys with different
flux limits in order to cover a wide range in quasar $B$-band
luminosity within the given redshift range.  Eight of the quasars are
from the faint Durham/AAT UVX survey of Boyle et al.\ (1990) and ten of 
the quasars are from the intermediate luminosity
Large Bright Quasar Survey by Hewett, Foltz
\& Chaffee (1995).  There are also two high-luminosity quasars in the
sample, selected from the Bright Quasar Survey (BQS) by
Schmidt \& Green (1983).
 
\begin{figure}
\epsfig{file=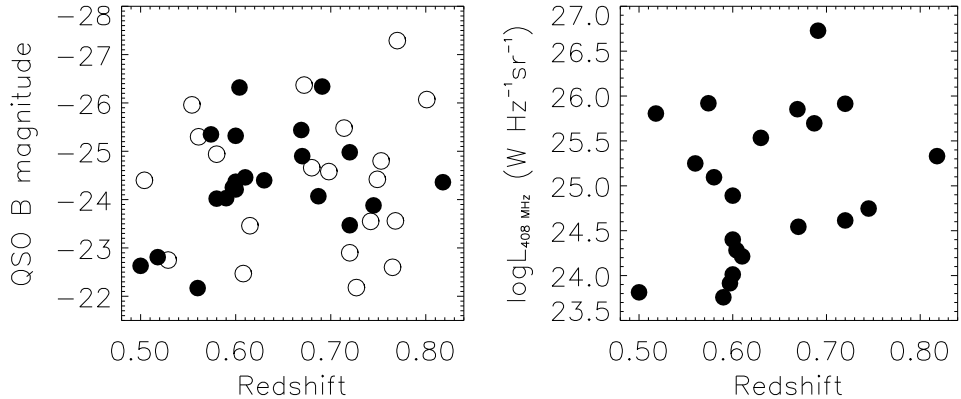}
\caption{The left-hand plot shows the location of the RLQs (filled
circles) and the RQQs (open circles) in the redshift-luminosity
plane. To the right are the radio luminosities at 408 MHz of the RLQs
as a function of redshift.}
\label{fig1}
\end{figure}

In Fig.~\ref{fig1} we show the distribution of the RLQs and the RQQs in
the redshift-luminosity plane. There is a slight tendency for the
quasars at bright $M_{B}$'s to appear at the highest redshifts. The correlation
between redshift and luminosity is a well-known feature in
flux-limited samples, but here it may instead be an artifact of the
narrow redshift range of the sample.  In our analysis we investigate
if there are any correlations between the environmental richness and
the quasar redshift and luminosity. It is therefore important that
there is no underlying correlation between redshift and luminosity in
the samples.  We use Spearman's partial rank correlation coefficients
for this analysis, allowing the correlation coefficient between two
variables (e.g.\ environmental richness and radio luminosity) to be
determined when holding the third variable constant (redshift).  For
more details about the RLQ sample, see Wold et al.\ (2000). The RQQ
sample will be presented by Wold et al.\ (in prep.).

\section{Control fields}

Since we are concerned with investigating if there is an excess of
galaxies in the quasar fields, we aim to have a good determination of
the foreground and background counts.  For this purpose, we obtained
several images of random fields in the sky at approximately the same
galactic latitudes as the quasar fields. There are twelve different
control fields of which five were imaged in two filters. In total,
they cover 18 arcmin$^2$ in $V$, 58.7 arcmin$^2$ in $R$ and 73.7
arcmin$^2$ in $I$.  They were imaged along with sources in the quasar
sample, so they have the same depth and were obtained in exactly the
same manner as the quasar fields. This is important in order to obtain
a robust estimate of the galaxy clustering in the quasar fields (Yee
\& L{\'o}pez-Cruz 1999).

\section{Observations}

Most of the data were obtained using the High Resolution Adaptive
Camera (HiRAC) at the NOT, equipped with either the 1k SiTe or the 2k
Loral CCD giving a field of view of 3$\times$3 and 3.7$\times$3.7
arcmin, respectively.  Some images were also obtained using the ALFOSC
in imaging mode, with the 2k Loral CCD with pixel scale 0.189 arcsec.
Typically, the integrations were divided into four exposures of 600 s 
each.  Two RLQ fields were imaged with the 107-in telescope at the
McDonald Observatory, and five RLQ fields with the HST (Serjeant,
Rawlings \& Lacy 1997).

The bulk of the data were obtained under photometric conditions, and
the seeing FWHM is less than one arcsec in 11 out of the 14 RLQ fields
imaged with the NOT, and in 15 out of the 20 RQQ fields. In several
images `fuzz' from the host galaxy is clearly visible around the
stellar image of the quasar.

The filters were chosen such as to give preference to early-type
galaxies with strong 4000 {\AA} breaks at the quasar redshifts.  For
an early-type galaxy at $z \geq 0.67$ the 4000 {\AA} break moves from
$R$-band into $I$-band, so for the $z \geq 0.67$ quasars we used
$I$-band imaging and for the $z < 0.67$ quasars we used $R$-band.  We
also imaged 20 of the quasar fields in two filters, either $V$ and
$R$, or $R$ and $I$ depending on the redshift of the quasar so as to
straddle the rest frame 4000 {\AA} break.

For photometry and object detection we processed the images in {\sc
focas} (Faint Object Classification and Analysis System).  By
performing completeness simulations in the images we find that the
data are complete down to 24.0 in $V$, 23.5 in $R$ and 23.0 in $I$,
with errors of $\pm$0.3 mag at the limits.  All quasar fields, except
two, lie at galactic latitudes $|b|>42^{\circ}$ and have galactic
reddening $E\left(B-V\right)<0.063$.  We corrected for galactic
extinction using an electronic version of the maps by Burstein \&
Heiles (1982) and the Galactic extinction law by Cardelli, Clayton \&
Mathis (1989).

\section{Results}

To quantify the amount of excess galaxies in the quasars fields we use
the amplitude, $B_{\rm gq}$, of the spatial galaxy--quasar
cross-correlation function, $\xi\left(r\right)=B_{\rm gq}r^{-\gamma}$,
where $\gamma=1.77$. The amplitude is evaluated at a fixed radius of
0.5 Mpc at the quasar redshift, corresponding to $\approx 1$ arcmin at
$z=0.7$ and has units of Mpc$^{1.77}$.  Longair \& Seldner (1979)
showed that $B_{\rm gq}$ can be found by first obtaining the amplitude
of the {\em angular} cross-correlation function, which is directly
proportional to the relative excess of galaxies. See Wold et al.\ (2000) for 
details of the analysis.

\begin{figure}
\epsfig{file=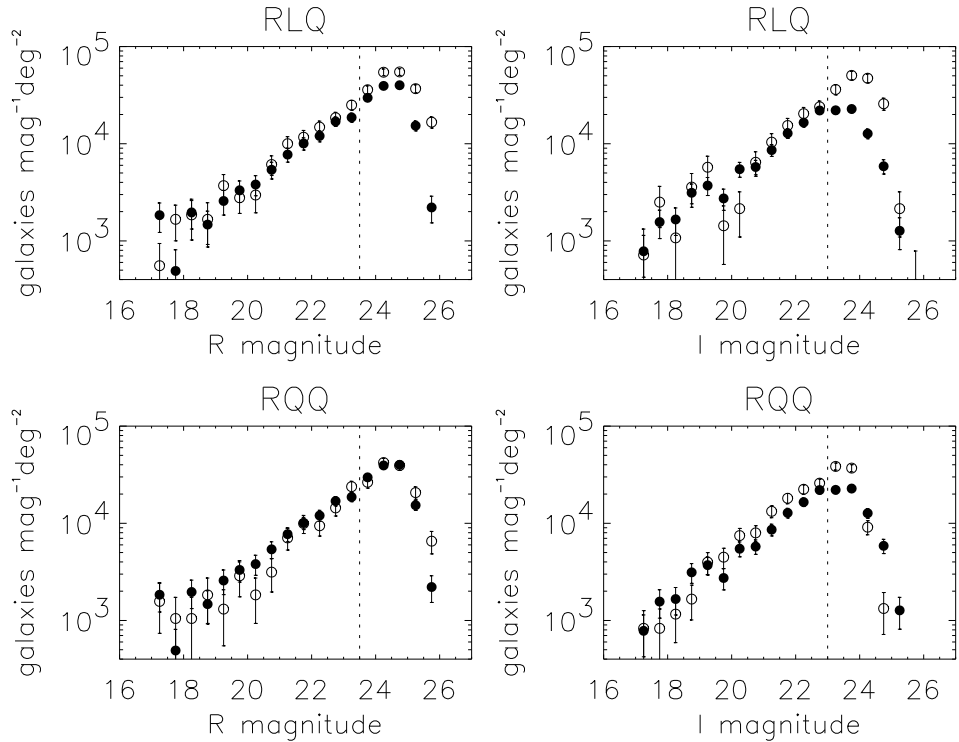}
\caption{Average galaxy number counts in the quasar (open circles) and
control fields (filled circles). The dotted lines show the
completeness limits of $R=23.5$ and $I=23.0$.}
\label{fig2}
\end{figure}

We counted the number of galaxies within the 0.5 Mpc radius in the
quasar fields and averaged the counts for the $R$ and $I$-band data,
i.e.\ for the $z<0.67$ and the $z \ge 0.67$ quasar fields. The average
counts are shown in Fig.~\ref{fig2} where we also have plotted the
average galaxy counts from the control images for comparison. The two
left plots in Fig.~\ref{fig2} show the average $R$-band counts for the
$z<0.67$ RLQ (13) and RQQ (7) fields, and the two plots to the right
show the average $I$-band counts for the $z \geq 0.67$ $I$-band RLQ
(6) and RQQ (13) fields.  In the RLQ fields there appears to be a small
excess of faint galaxies at $R$ and $I \gsim 21$, whereas the
$R$-band RQQ fields show no excess. However,
there is a clear excess of galaxies in the $z \geq 0.67$ RQQ fields at
$I>20$.  The errors in the background galaxy counts were calculated as
1.3$\sqrt{N}$ in order to take into account the non-random
fluctuations in the counts due to the clustered nature of field
galaxies.

\begin{figure}
\epsfig{file=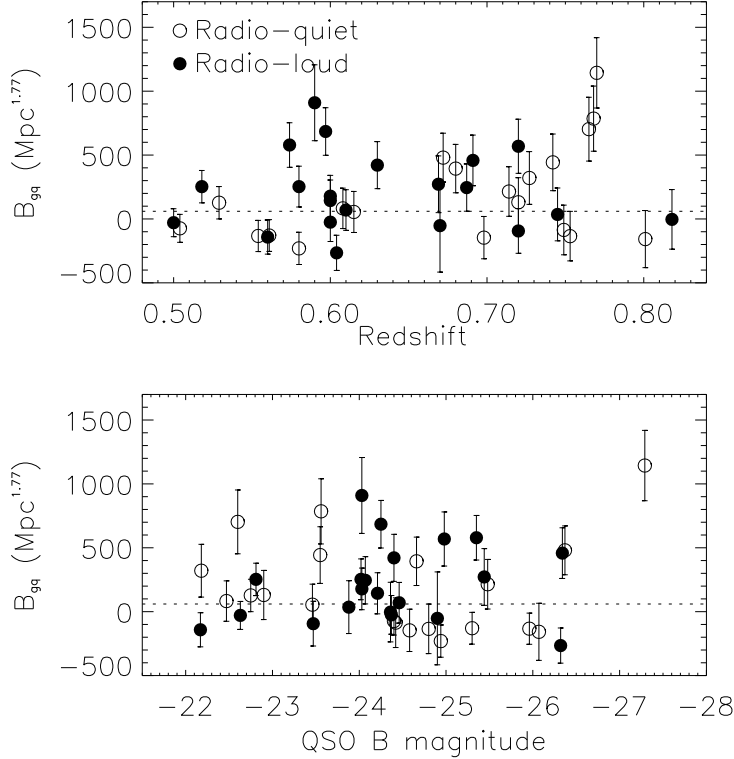}
\caption{Clustering amplitudes for the RLQs (filled circles) and the
RQQs (open circles) as a function of redshift and quasar $B$ absolute
magnitude. The dotted lines show the value of the local galaxy-galaxy
clustering amplitude $B_{gg}=60$ Mpc$^{1.77}$ (Davis \& Peebles
1983). Note that the environmental richness is similar for both RLQs
and RQQs.}
\label{fig3}
\end{figure}

We calculated the net excess of galaxies in each quasar field by
subtracting the average background counts, and thereafter computed the
clustering amplitude, $B_{\rm gq}$.  In Fig.~\ref{fig3} we show
$B_{\rm gq}$ for the RLQ and the RQQ fields as a function of redshift
and quasar $B$ absolute magnitude. The dotted line across the plots
show the value obtained by Davis \& Peebles (1983) for the amplitude
of the local galaxy-galaxy auto-correlation function, $B_{\rm gg}=60$
Mpc$^{1.77}$. The mean clustering amplitudes for the RLQ and the RQQ
samples are 213$\pm$66 and 189$\pm$83 Mpc$^{1.77}$, respectively.  We
thus make two observations, first that the mean clustering amplitudes for
the quasar fields are significantly larger than that of local
galaxies, implying that the quasars exist in fields with richer than
average galaxy density.  Second, we note that the mean clustering
amplitudes for the two samples are practically indistinguishable,
i.e.\ {\em on average, there is no difference in the galaxy
environments on 0.5 Mpc scales for the RLQs and the RQQs}. On average,
both the RLQs and the RQQs seem to prefer environments similar to galaxy
groups and poorer clusters of galaxies.

\begin{figure}
\epsfig{file=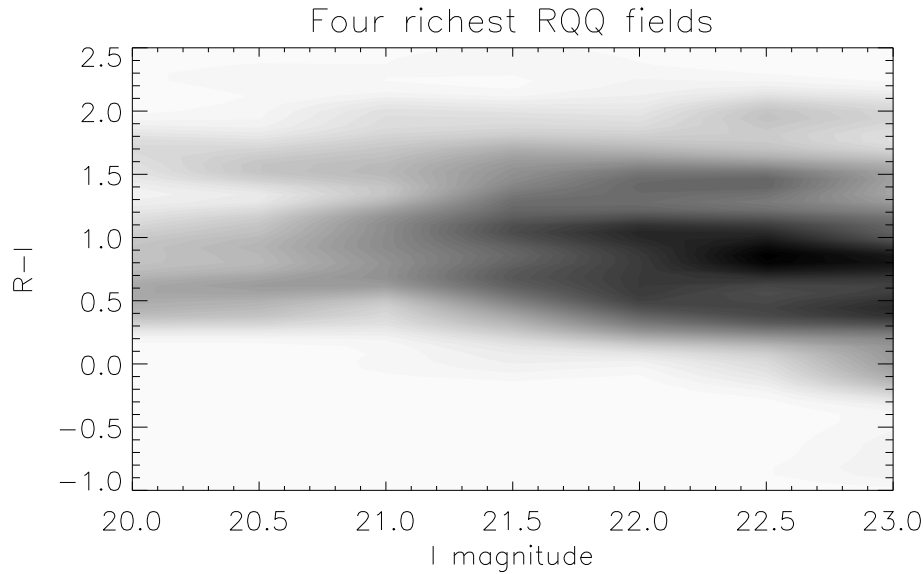}
\caption{Colour-magnitude diagram of the galaxies in the four richest
RQQ fields ($\left<z\right>=0.74\pm0.02$). There are 364 galaxies
in this plot, and the distribution is
smoothed with a Gaussian filter with smoothing lengths 0.5 mag in $I$
magnitude and 0.125 in $R-I$. Few galaxy clusters are known at $z
\gsim 0.7$, but it seems that the expected colour of the red sequence
in clusters at these redshifts lies in the range $1.3 \lsim R-I \lsim
2$ (Clowe et al.\ 1998; Luppino \& Kaiser 1997).}
\label{fig4}
\end{figure}

The quasar environments span a wide range, however. Some individual
fields show no significant excess of galaxies, and other 
fields appear to be very rich, e.g.\ the field around the RQQ BQS
1538$+$477 with $B_{\rm gq}$ in the range 1100--1200 Mpc$^{1.77}$ and
a galaxy excess of $\approx 40$--50 galaxies.  Wold et al.\ (2000)
argue that an amplitude of $\approx 740$ Mpc$^{1.77}$ corresponds to
Abell richness class $\gsim$ 1, so the cluster candidate around BQS
1538$+$447 must qualify as a richness class 2 cluster.  Two other
fields with clustering amplitudes of 785$\pm$255 and 703$\pm$250
Mpc$^{1.77}$ are probable Abell class 1 clusters.
Plotting the galaxies in the four richest RQQ fields in a
colour-magnitude diagram, reveals a hint of a red sequence at $R-I
\approx 1.5$--1.6, i.e.\ tentative evidence that these fields likely
contain galaxy clusters at $z\sim 0.7$--0.8, see Fig.~\ref{fig4}.

As seen in Fig.~\ref{fig3} there are no obvious trends in $B_{\rm gq}$
with either redshift or quasar luminosity. There is a hint that the
low-$z$ RQQ fields have lower clustering amplitude than the higher
redshift RQQ fields, but this is most likely an artifact of the narrow
redshift range. The Spearman partial rank correlation coefficient
giving the correlation between $B_{\rm gq}$ and redshift is 0.4 with a
1.5$\sigma$ significance.

\section{A link between radio luminosity and environmental richness?}

\begin{figure}
\epsfig{file=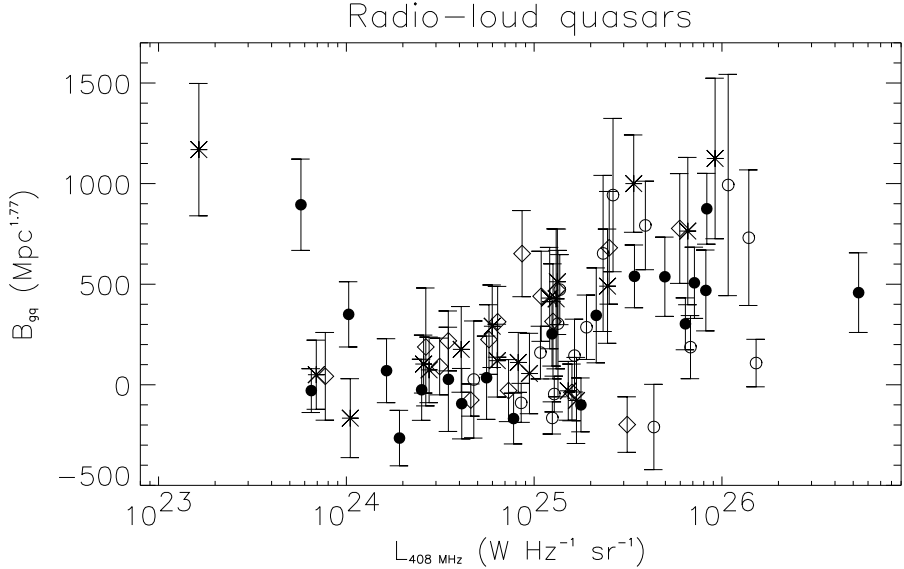}
\caption{Clustering amplitude for the quasars in our sample (filled
circles) plotted together with RLQs in the literature as a function of
radio luminosity. Open circles show data from Yee \& Green (1987),
stars are data from Ellingson et al.\ (1991b) and diamonds are data
from Yee \& Ellingson (1993). We find a correlation coefficient of 0.4
with a 3.4$\sigma$ significance.}
\label{fig5}
\end{figure}

Looking at $B_{\rm gq}$ for the RLQs as a function of radio luminosity,
reveals a weak, but significant, correlation between $B_{\rm gq}$ and
radio luminosity with much scatter.  This is shown in Fig.~\ref{fig5}
where we also have plotted $B_{\rm gq}$'s for RLQ fields as found by
Yee \& Green (1987), Ellingson et al.\ (1991b) and Yee \& Ellingson
(1993).  Here the, correlation coefficient between $B_{\rm gq}$ and
$L_{\rm 408 MHz}$, holding redshift constant, is 0.4 with a 3.4$\sigma$
significance.

Does the correlation between radio luminosity and environmental
richness imply that environment is the primary factor in controlling
the radio luminosity of a RLQ?  There are at least three ways in which
such a situation could come about. In the first, the environment
determines the bulk kinetic power in the radio jets. According to
the relation between a galaxy's black hole mass and the mass of the
spheroidal component (Kormendy \& Richstone 1995; Magorrian et al.\
1998), giant elliptical hosts of radio galaxies and RLQs should have
high black hole masses, $\sim 10^{8}$--$10^{9}$ M$_{\odot}$. Assuming
that the radio jets are powered by accretion and that the accretion
rate is proportional to the black hole mass, galaxies with massive
black holes will power more luminous radio sources.  These massive
galaxies will prefer richer environments and thus the correlation
between radio luminosity and environmental richness may just reflect
an increasing mass of the host.

The second possibility is that more fuel for the quasar and its
radio jets is available in a group or cluster environment.  A group or a
poor cluster environment may be ideal for the fuelling of a black hole
as encounters will be more common than in the field and will be of low
enough relative velocity to disrupt the interacting galaxies and cause
gas to flow into the centre (Ellingson et al.\ 1991a).

A third possibility is that the radio luminosity is almost
independent of the bulk kinetic power in the radio jets and is instead
largely determined by the density of the environment into which the
source expands. Wold et al.\ (2000) constructed a simple model with
these assumptions and found that the predicted relation between
$B_{\rm gq}$ and radio luminosity was much too steep to fit the data,
thereby ruling out as strong a $B_{\rm gq}$--$L_{\rm 408 MHz}$
dependence as we would see if all RLQs had the same jet power and
environment was entirely responsible for determining the radio
luminosity. Nevertheless, as the luminosity function for the radio jet
power is likely to be steeply declining at high powers, it seems not
unlikely that selection effects could operate to produce some
correlation between $L_{\rm 408 MHz}$ and $B_{\rm gq}$ without it
being as strong as it would be in this rather extreme model in which
the jet power is the same for all sources.

Given the large scatter in the correlation it is however quite
possible that both environment and radio jet power play important
roles in determining the radio luminosity. The relationship between
radio sources and their environments must be complex, and the vast
majority of radio sources may lie in some sort of cluster-like
environment, from groups of only a few galaxies to clusters as rich as
Abell class 1 or more.

Given this correlation, one might expect the RQQs, with radio
luminosities typically two-three orders of magnitude lower than the
RLQs, to be sited in poorer environments. Instead we find
that the environments around RLQs and RQQs are indistinguishable.  This
is however fully consistent with the $B_{\rm gq}$-radio luminosity
correlation.  That the large-scale environments around RLQs and RQQs are
similar suggests that the process that decides on radio loudness in a
quasar is not dependent on the environment on Mpc scales, but may
instead be found in the central regions of the host galaxy.

When an epoch of quasar activity is triggered in a galaxy, e.g.\ as a
result of galaxy interactions and mergers, then some central process
decides whether it becomes a RLQ or a RQQ.  If a RLQ is born, the
environment into which the source expands will to some extent
determine its radio luminosity, and we observe the $B_{\rm
gq}$-$L_{\rm 408 MHz}$ correlation.  If instead a RQQ is born, we do
not observe any correlations between quasar luminosity (in this case
optical luminosity) and environmental richness, because this quasar
does not have extended radio lobes that can interact with the
surrounding galaxy environment.  The radio jets and the radio lobes
extending beyond the host galaxy thus work like sensors from which we
can read off the physical conditions in the intergalactic medium.

\section{The evolutionary states of AGN-selected clusters}

The fuelling mechanism of luminous AGN is still not understood,
although it is thought that companion galaxies in groups and clusters
may be able to supply fuel either through mergers or via a cluster
cooling flow (e.g.\ Hall, Ellingson \& Green 1997).  The merger and
cooling flow models, however, make very different predictions for the
evolutionary state of the cluster surrounding the AGN. If the AGN is
fuelled by mergers and interactions, we might expect that the cluster
is still forming by merging of sub-clumps, but in the cooling flow
scenario the cluster may be well-established and virialized.  We have
therefore started to investigate the evolutionary states of AGN-selected galaxy
clusters by weak lensing techniques.

Using deep images obtained with the ALFOSC on the NOT in sub-arcsec
seeing, we have mapped the projected mass distribution in rich, X-ray luminous
AGN-selected clusters, and preliminary results show that we have comfortable
detections of the weak lensing signal ($>3\sigma$).  These
observations will also allow us to estimate the mass-to-light ratio
for the clusters, and to investigate whether the mass-to-light ratio
is different than for clusters selected solely on the basis of bright
optical or X-ray emission from their baryonic matter component.

The weak lensing technique is being used more and more frequently
since it is a powerful method for investigating galaxy clusters (see
e.g.\ Dahle et al., these proceedings). The combination of the good
seeing conditions at the NOT and the wide field imagers with good
resolution currently available and underway at the NOT (the
focalreducer FRED) makes the NOT a powerful instrument for doing weak
lensing observations.

\acknowledgments

We are grateful to the staff at the NOT and the McDonald Observatory
for help with the observations. We also thank the British
Research Council and the Research Council of Norway for support. MW
also wishes to thank the organizers of the meeting for support.

\end{document}